\documentclass[aps,pre,preprint,groupedaddress]{revtex4-1}

\usepackage{graphicx}
\usepackage{siunitx}
\usepackage[version=3]{mhchem}
\usepackage{subfig}
\usepackage{rotating}
\usepackage{colortbl}
\usepackage{hyperref}

\begin{document}


\title{Time-dependent wave selection for information processing in excitable media}


\author{William M. Stevens}
\email{william.stevens@uwe.ac.uk}
\affiliation{Faculty of Environment and Technology, University of the West of England, Bristol, UK, BS16 1QY}
\author{Ishrat Jahan}
\affiliation{School of Life Sciences, University of the West of England, Bristol, UK, BS16 1QY}
\author{Ben de Lacy Costello}
\affiliation{School of Life Sciences, University of the West of England, Bristol, UK, BS16 1QY}
\author{Andrew Adamatzky}
\affiliation{Faculty of Environment and Technology, University of the West of England, Bristol, UK, BS16 1QY}


\date{\today}

\begin{abstract}
We demonstrate an improved technique for implementing logic circuits in
light-sensitive chemical excitable media. The technique makes use of the constant-speed propagation of waves
along defined channels in an excitable medium based on the Belousov-Zhabotinsky reaction, along with the mutual annihilation of colliding waves. What distinguishes this work from previous work in this area is that  
regions where channels meet at a junction can periodically alternate between permitting
the propagation of waves and blocking them. These valve-like areas are
used to select waves based on the length of time that it takes waves to propagate from one valve
to another.
In an experimental implementation, the channels which make up the circuit layout are projected by a digital projector
connected to a computer. Excitable channels are projected as dark areas, unexcitable regions as light areas.
Valves alternate between dark and light: every valve has the same period and phase, with a 50\% duty cycle.
This scheme can be used to make logic gates based on combinations of OR and AND-NOT operations, with few geometrical constraints.
Because there are few geometrical constraints, compact circuits can be implemented.
Experimental results from an implementation of a 4-bit input, 2-bit output integer square root circuit are given. This is the most complex logic circuit that has been implemented in BZ excitable media to date.

\end{abstract}


\maketitle

\section{Introduction \label{introduction}}

The use of discrete propagating waves for information processing with Belousov-Zhabotinsky (BZ) excitable media began with the capillary tube logic gates constructed by T\'{o}th and Showalter \cite{toth1995}. Since then many different schemes for information processing in BZ excitable media have been devised \cite{adamatzky2005}.
Steinbock et. al. implemented a logic scheme in which a catalyst for the BZ reaction is printed onto regions of a planar medium \cite{steinbock1996} -- waves are able to propagate in these regions, but not elsewhere, and the operation performed by a region is determined by the geometry of the region.
Agladze et. al. \cite{agladze1996} were the first to carry out computer simulations and experimental demonstrations of a `diode' geometrical arrangement which permits a wave to travel in one direction but not in the opposite direction. Other diode structures were reported in \cite{toth2001} and \cite{gorecka2007}.
Motoike and Yoshikawa \cite{motoike1999} simulated several logic circuit elements including AND, OR and NOT gates, a time difference detector, and a memory element. An experimental implementation of a similar memory element was reported in \cite{motoike2001}.
Sielewiesiuk and G\'{o}recki simulated a cross junction capable of acting as a coincidence detector and a switch \cite{sielewiesiuk2001}, and also showed that when a train of pulses passes from one excitable region to another over a passive barrier, pulses will sometimes `dropout' of the train. The rate of dropout is determined by many different parameters \cite{sielewiesiuk2002}.
 
The use of a light sensitive Ruthenium catalyst greatly simplifies the process of applying a pattern to 
a planar BZ excitable medium. The entire medium can be impregnated with the catalyst, but the reaction can be suppressed in regions where it is not wanted by illuminating those regions \cite{kadar1997}. Typically a slide projector or a digital projector is used for this \cite{kuhnert1989,gorecki2003,ichino2003}.
G\'{o}recki et. al. \cite{gorecki2003} designed unary and binary counting circuits, able to count the number of waves entering a channel. All of the components of the counters were tested in experiments, although the complete counters were not. Ichino et. al. \cite{ichino2003} showed that the behaviour of a diode structure and a time difference discriminator can be altered by altering the level of illumination. 

Whereas the construction of basic logical elements in BZ excitable media depends upon finding a mapping between logical operations and the behaviour of waves in a medium, 
several works have demonstrated that some more complex information processing tasks can be carried out by finding a good mapping or analogy between the task and the dynamics of the medium.
Kuhnert et. al. projected an image directly onto a light-sensitive medium to perform an image processing task  \cite{kuhnert1989}. Steinbock et. al. used circularly spreading waves to help find the shortest path through labyrinths \cite{steinbock1995}, where the interior of the labyrinth was excitable but
the labyrinth walls and exterior of the labyrinth were not.
G\'{o}recka and G\'{o}recki \cite{gorecka2003} simulated the operation of a bandpass filter, capable of outputting a signal whenever the time interval separating the arrival of two input pulses matched a constant value determined by the dimensions of the filter.
Motoike and Yoshikawa \cite{motoike2003} simulated a wave train comparator and subtractor which operate directly on trains of waves.
Nagahara et. el. \cite{nagahara2004} demonstrated a direction detector, capable of detecting the direction of origin of a spreading circular wave.
G\'{o}recki et. al. \cite{gorecki2005} showed how an array of excitable paths can be used to make a distance detector by exploiting the phenomenon described in \cite{sielewiesiuk2002}. G\'{o}recka and G\'{o}recki \cite{gorecka2006} simulated a device analogous to an integrate-and-fire neuron and showed that it can effect a four input Boolean threshold function much more simply than can be done using AND and OR gates. 

Adamatzky et. al. explored the potential of BZ excitable media as a substrate for implementing architecture-less collision based computing schemes \cite{adamatzky2004}. In this paradigm, exemplified in a cellular automaton model by Margolus in \cite{margolus2002}, the substrate does not contain any information about the computation to be carried out: it contains no structure that determines a specific function. This information is contained in the pattern and timing of events that are applied to the substrate. For certain levels of illumination, light sensitive BZ excitable media have a `subexcitable' mode in which waves of a critical size travel forward and may either expand or contract under the influence of small perturbations \cite{mihaliuk2002}. Waves below this size slowly contract as they move forward. Waves above this size slowly expand. Collisions between subexcitable waves can be interpreted as logical operations \cite{delacycostello2005}, \cite{toth2009}. Subexcitability combined with structured media was used in \cite{delacycostello2011}. In \cite{adamatzky2011,holley2011}, interactions between subexcitable waves were contained within circular discs to help control their behaviour. 
Spiral waves were used in \cite{delacycostello2009} as the basis for a glider-gun-like mechanism, capable of producing a continuous stream of discrete waves.

The variety of phenomena that are available in BZ excitable media and which have been used for constructing information
processing structures justifies the use
of these media for exploring new ways of making information processing mechanisms. By investigating all of the possible ways in which these phenomena may be used for
information processing, we equip ourselves with a toolset which can be applied when we encounter new media
that share some of these phenomena. 

The work in this paper has several things in common with the work of Steinbock et. al \cite{steinbock1996}. It makes use of the same
two phenomena: the constant speed propagation of waves in BZ excitable media, and the mutual annihilation of two wavefronts that meet.
Differences in wave propagation time between
paths from the inputs to the outputs of a region of excitable medium are exploited for implementing logical
operations. Where this work differs from that of Steinbock et. al. is that the region through which waves
propagate is not a fixed pattern, but contains areas that alternate between being excitable and unexcitable in a simple periodic way. This is possible because the region is defined by a pattern projected
using a digital projector. The ability to periodically switch areas from excitable to unexcitable and back is used to overcome
one of the limitations of the work of Steinbock et. al, described below. 

\begin{figure}
\includegraphics[width=2.0in]{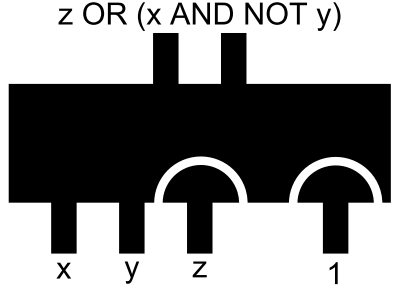}
\caption{The $z$ OR ($x$ AND NOT $y$) gate from \cite{steinbock1996} executing the case $z=1,x=0,y=0$.\label{steinbock}}
\end{figure}

Steinbock et. al. use the presence or absence of a wave in an input channel to represent Boolean 1 and 0 values at the inputs to a gate. All inputs to a gate must be applied at the same time. At the output of a gate, the synchrony or asynchrony of the arrival of waves in a pair of
output channels is used to represent Boolean 1 and 0 respectively: if the waves arrive at both output channels at the same time, this represents 1, if the waves 
arrive at different times, this represents 0.  One of the limitations of the gates of Steinbock et. al. is that the timing of the output from a gate is dependent on the values of the inputs. Consequently, when gates are cascaded there is no way
to be sure that all inputs at a gate downstream in the cascade arrive at the same time. This is illustrated in
Figure \ref{steinbock}, which shows the $z$ OR ($x$ AND NOT $y$) gate from \cite{steinbock1996}. In this gate, if
$z=1$, a 1 output will emerge from the gate after the time that it takes a wave to propagate from $z$ to
the output. However, for
the case $z=0, x=1, y=0$, a 1 output will emerge from the gate after the time that it takes a wave
to propagate from $x$ to the output, which is different from the case when $z=1$.
Circuits implemented using the scheme described in this paper do not have this limitation.
Outputs emerge from a circuit at a constant time after the inputs are applied.

\section{Time-dependent wave selection \label{interacting_waves}}

\subsection{Fork structures}

\begin{figure}
\includegraphics[width=2.5in]{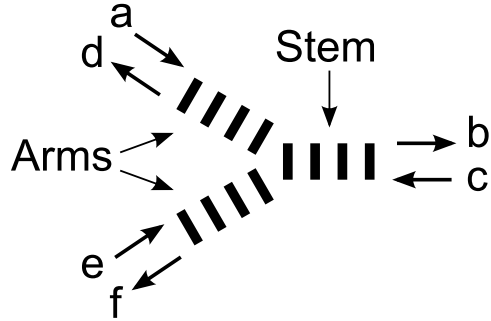}
\caption{A domino fork. Letters denote events. Input events ($a$, $c$, $e$) correspond to a domino toppling into the fork, output events ($b$, $d$, $f$) correspond to a domino toppling out of the fork.\label{dominofork}}
\end{figure}

The path which led to the work in this paper began by observing the similarity between the behaviour of toppling dominoes and the behaviour of waves in channels of BZ excitable media. Domino fork structures can be used as the basis for making Boolean logic circuits \cite{okeefe2009,stevens2011b}. A domino fork structure, shown in Figure \ref{dominofork}, has three lines of dominoes meeting at a junction. We call two of these lines arms, and the third we call the stem.
The end of each line is labelled with two different events: one event corresponds to a domino toppling into the structure, the other event corresponds to a domino toppling out of the structure. The behaviour of the domino fork can be described in terms of the relationships between these events. Informally, we can say that
either event $a$ or event $e$ (or both) will cause event $b$ (unless event $c$ occurs in the mean time), and that event $c$ will cause events $d$ and $f$ (unless $a$ or $e$ occur in the mean time).

\begin{figure}
  \subfloat[Splitting at a fork junction in a simulation of subexcitable BZ-reaction media.]{\label{bzfork1}\includegraphics[width=1.5in]{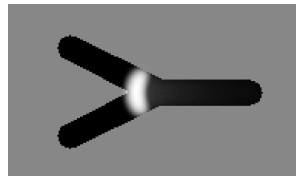}}   
  \hspace{6mm}
  \subfloat[Directional propagation at a fork junction in a simulation of subexcitable BZ-reaction media.]{\label{bzfork2}\includegraphics[width=1.5in]{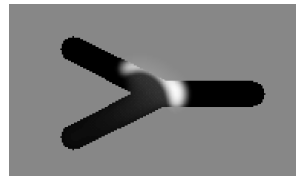}}   
  \caption{A simulation of a fork structure in BZ excitable media.\label{bzfork}}  
\end{figure}

An important feature of fork behaviour is that an event entering the fork from one arm will not influence the the other arm; it will only cause an event to emerge from the stem of the fork.
In the domino fork structure this behaviour is obtained through directionally-dependent propagation of a toppling domino wave: a domino can only influence other dominoes if they stand in the direction in which it topples. A fork structure with behaviour based on directionally-dependent wave propagation can also be implemented in BZ excitable media. A structure similar to this can be found in \cite{motoike1999} and \cite{delacycostello2011}, and Figure \ref{bzfork} shows the result of simulating a fork structure which depends upon subexcitability (as defined in section \ref{introduction}) for its operation. Subexcitability is used to prevent a wave that is propagating along one arm from entering the other arm. The obvious difference between a domino fork and a fork implemented in BZ excitable media is that whereas the domino fork cannot be used again once dominoes have toppled, the excitable medium fork can be used again and again. 

\begin{figure}
  \subfloat[Splitting at a fork junction in subexcitable BZ-reaction media (compare with Figure \ref{bzfork1}).]{\label{expfork1}\includegraphics[width=1.5in]{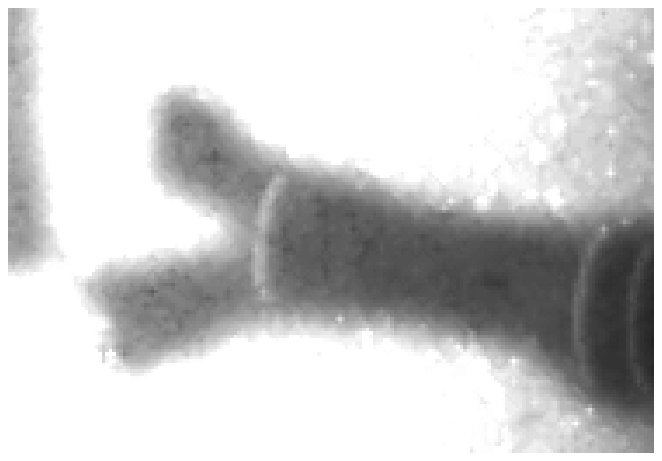}}   
  \hspace{6mm}
  \subfloat[Directional propagation at a fork junction in subexcitable BZ-reaction media (compare with Figure \ref{bzfork2}).]{\label{expfork2}\includegraphics[width=1.5in]{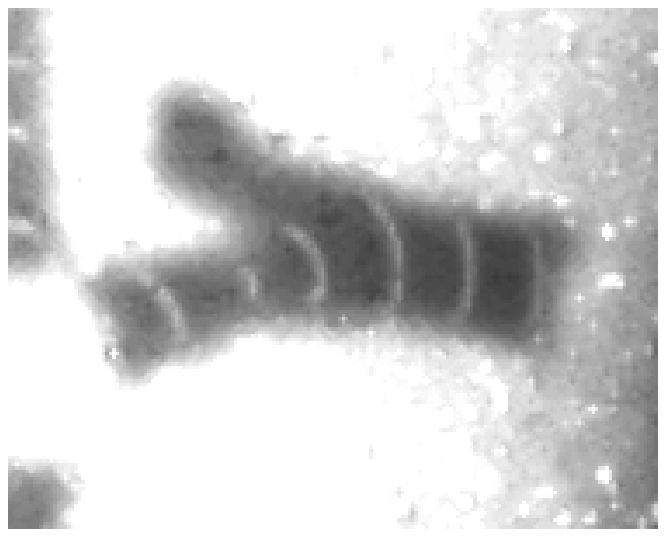}}   
  \caption{Experimental implementation of a fork structure in BZ excitable media.\label{expfork}}  
\end{figure}

Simulations of excitable media circuits were able to reproduce some of the behaviour described in \cite{stevens2011b}. The success of these simulations led to attempts to implement them experimentally. Figure \ref{expfork} shows snapshots from an experiment exhibiting the same behaviour as shown in Figure \ref{bzfork}. 

Two problems were encountered in the course of conducting these experiments. The first was that the range of illumination levels in which the subexcitable regime can be obtained is both very narrow, and also sensitive to experimental conditions. If the illumination level is too high, the medium is unexcitable and waves die off after a short distance. If the illumination level is too low, the medium is fully excitable and directionally-dependent propagation cannot be obtained. The subexcitable illumination range was found to change from one location in the medium to another, and also to change over time during an experimental session. It was therefore difficult to obtain a subexcitable regime across a large region for a long period of time.

The second problem was that harnessing subexcitability to obtain fork behaviour places constraints on the geometry of a circuit made from forks: the arms of the fork must meet the stem within a certain range of angles, and must extend for a certain distance away from the fork junction before they can turn a corner. 
These constraints limit the complexity of circuits that can fit into the area available for experiments (about 5cm by 5cm). 
 
\subsection{Selecting waves based on their propagation time}

These problems led to a search for an alternative way to obtain the same behaviour.
Rather than use subexcitability to prevent a wave that is travelling along one arm from entering the other arm, we instead arrange the fork so that the time that it takes for a wave to propagate from arm to stem (or from stem to arm) is twice as long as the time that it takes for a wave to propagate from arm to arm. We then make sure that only waves taking the longer time are permitted to exit the fork.
 
\begin{figure}
\includegraphics[width=1.5in]{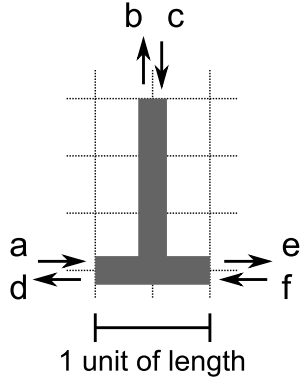}
\caption{A T-shaped region of excitable medium. The region is fully excitable, so a wave entering one terminal will propagate throughout the region. A unit of length is shown. A unit of time is defined as the time that it takes a wave to propagate one unit of length.\label{tshapedregion}}
\end{figure}

Consider the T-shaped region of excitable medium shown in Figure \ref{tshapedregion}. This region is fully excitable: a wave entering the region will propagate throughout the region. A unit of length is defined in
Figure \ref{tshapedregion} and we define a unit of time to be the time that it takes a wave to propagate one
unit of length.
There are three terminals at the boundary of this region at which
waves can be sent into the region, or observed emerging from the region. We call the passage of a wave in a 
particular direction through one of these terminals an event, and we label events using letters.

Table \ref{tshapebehaviour} shows which events emerge from the region at times $t_0+1$ and $t_0+2$ for
all possible combinations of input events $a$, $c$ and $f$ at time $t_0$.
If we let an event at a specified time represent a Boolean 1 value and
the absence of an event at a specified time represent a Boolean 0 value, then
we can see from this table that if we ignore or somehow block events that occur at $t_0+1$ and only pay attention to events
that occur at $t_0$ and $t_0+2$ then this region can be used to implement two different Boolean operations:

\begin{itemize}
\item[] If $c$ AND-NOT ($a$ OR $f$) occur at $t_0$ then both $e$ and $d$ will occur at $t_0+2$
\item[] If ($a$ OR $f$) AND-NOT $c$ occur at $t_0$ then $b$ will occur at $t_0+2$
\end{itemize}

\begin{table}
\caption{The behaviour of the region in Figure \ref{tshapedregion} for all possible combinations of input events $a$, $c$ and $f$ at time $t_0$.\label{tshapebehaviour}}
\begin{ruledtabular}
\begin{tabular}{|l|l|l|}
\hline
Events at $t_0$ & Events at $t_0+1$ & Events at $t_0+2$ \\
\hline
None & None & None \\
$a$ & $e$ & $b$ \\
$c$ & None & $d,e$ \\
$f$ & $d$ & $b$ \\
$a,c$ & $e$ & None \\
$a,f$ & None & $b$ \\
$c,f$ & $d$ & None \\
$a,c,f$ & None & None \\
\hline
\end{tabular}
\end{ruledtabular}
\end{table}

The main insight that led to the results in this paper is that 
events that occur at $t_0+1$ can be blocked by placing `valves' at each terminal, and ensuring that
the valves permit the passage of waves at $t_0$ and $t_0+2$, but do not permit the passage of waves at $t_0+1$.
In section \ref{experimental} we will see that in an experimental setup where the circuit layout is projected by a digital projector
onto the surface of a light-sensitive excitable medium, valves can be implemented by having valve regions of the projected circuit
switch between dark and light periodically: when the valve regions are dark they permit the passage of waves, when
they are light they inhibit the propagation of waves. We call this scheme time-dependent wave selection. 

\begin{figure}
\includegraphics[width=3.0in]{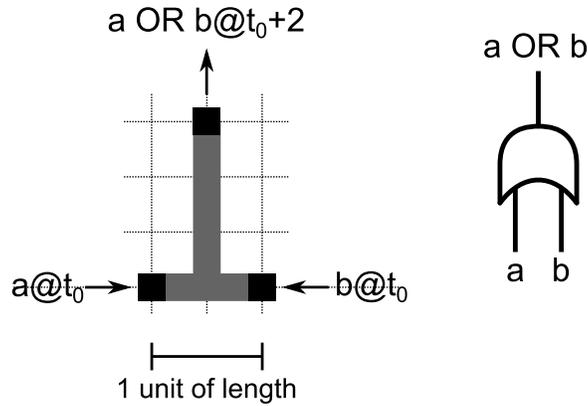}
\caption{A T-shaped region of excitable medium with valves at its terminals. This region implements a two input OR-gate.\label{tshapedregion_valves}}
\end{figure}

Figure \ref{tshapedregion_valves} illustrates this for a T-shaped region in which two terminals have been designated
as inputs $a$ and $b$ and the third terminal has been designated as an output. The output terminal outputs the function $a \textrm{ OR } b$ two time units after the inputs are applied.
In this and future figures, a location labelled using the notation $a@t_0$
means that a Boolean 1 value at that location is represented by the occurrence of event $a$ at time $t_0$, 
and that a Boolean 0 value at that location is represented by the absence of event $a$ at $t_0$. 

In order to permit some flexibility in the placement of valves, and in order to tolerate minor deviations
from perfect timing in simulations and experiments, valves alternate between being open (i.e. they permit the passage of waves) for a whole unit of time, then closed for a whole unit of time. Valves are open at $t_0+2n - 0.5 < t \le t_0+2n + 0.5$ for any integer $n$ and closed at other times.

\begin{figure}
\includegraphics[width=5.0in]{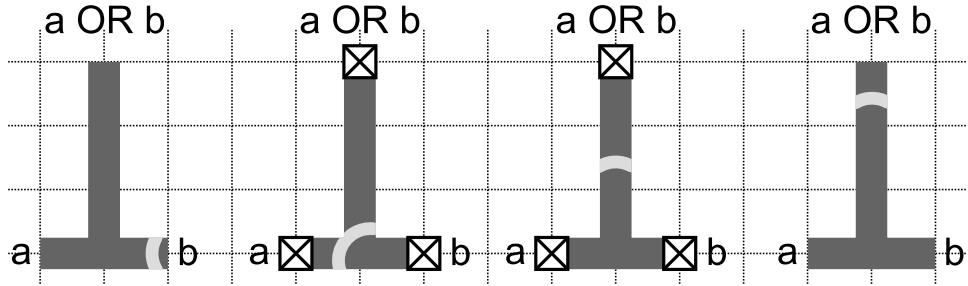}
\caption{A wave propagating from $b$ to the $a$ OR $b$ output. Squares with crosses inside represent closed valves. The left-hand case in this figure shows the wave shortly after $t_0$. Subsequent cases show the progress of the wave every time unit until shortly after $t_0+3$.  The wave never reaches $a$ because it is blocked by a closed valve.\label{prop_example}}
\end{figure}

Figure \ref{prop_example} illustrates what happens to a wave that enters $b$ at time $t_0$. The wave propagates throughout the excitable region, but it is unable to reach $a$ because valves are closed around the time that the wave reaches $a$. However, by the time the wave reaches the $a$ OR $b$ output, valves are open and the wave is not blocked.

Fork behaviour obtained through time-dependent wave selection overcomes the two problems that were encountered with trying to obtain the same behaviour using subexcitability. Time-dependent wave selection uses the fully excitable regime, and so it is not sensitive to experimental conditions. Additionally, it does not require
any particular geometry -- T-shapes can
be deformed in any way that preserves the wave propagation time between any pair of terminals.

Because time-dependendent wave selection does not place many constraints on the geometry of a structure, it is easy to make structures having more than three terminals, so the behaviour exhibited by the fork can be  extended.

\begin{figure}
\includegraphics[width=4.0in]{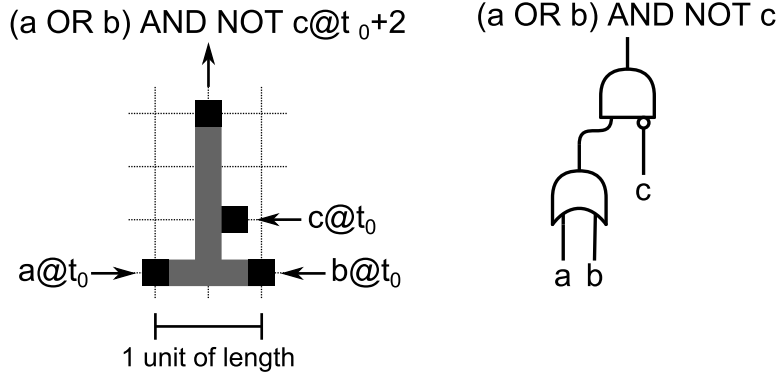}
\caption{A T-shaped region of excitable medium with an additional input. This structure implements the function ($a$ OR $b$) AND NOT $c$.\label{region2_valves}}
\end{figure}

The structure shown in Figure \ref{region2_valves} extends the OR-gate function from Figure \ref{tshapedregion_valves} by introducing an additional input, labelled $c$. Because this input is a unit distance away from all other terminals, a wave entering $c$ at $t_0$ will not emerge from any other terminal. However, it will annihilate any wave that enters at $a$ or $b$ at $t_0$, and thus prevent waves from $a$ or $b$ from reaching the output. This structure implements the function ($a$ OR $b$) AND NOT $c$.

\begin{figure}
\includegraphics[width=5.0in]{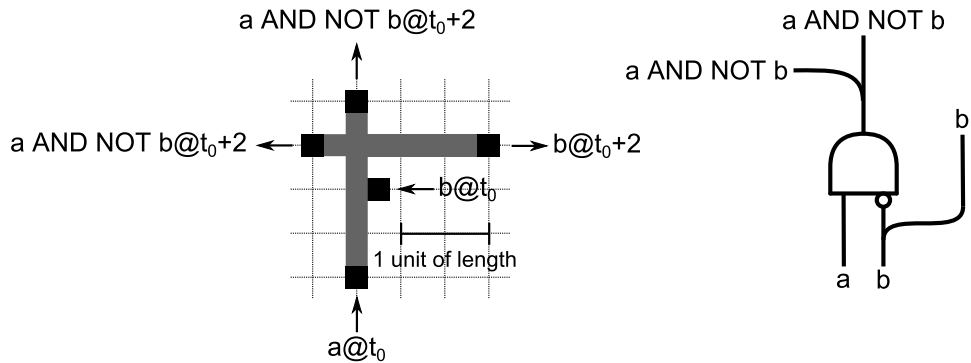}
\caption{A region of excitable medium that implements the function $a$ AND NOT $b$. The $a$ AND NOT $b$ output emerges from two different terminals, and the $b$ input is also made available at an output.\label{region3_valves}}
\end{figure}

The structure shown in Figure \ref{region3_valves} has five terminals: two inputs and three outputs. This structure implements the function $a$ AND NOT $b$, where the output emerges in two different locations, and the $b$ input is also made available at an output.

\section{A 4-bit square root circuit}

\begin{figure}
\includegraphics[width=5.5in]{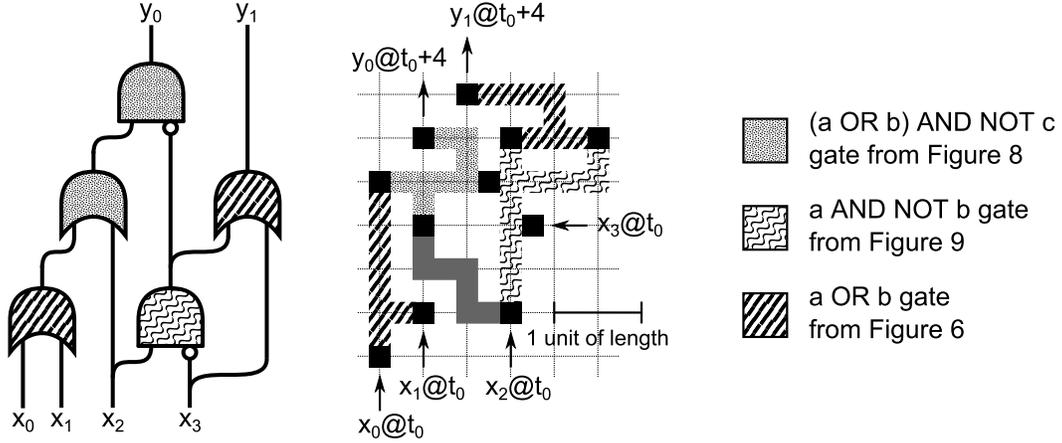}
\caption{A 4-bit integer square root circuit. $x_n$ is the $n^{th}$ binary digit of $x$ and $y_n$ is the $n^{th}$ binary digit of $y$. The logic gate symbols and excitable media channels are shaded to show the correspondence between the two. The key refers to the previous Figures in which the excitable media structures that are used here were introduced. \label{root4bit_sidebyside}}
\end{figure}

The left-hand side of Figure \ref{root4bit_sidebyside} shows a schematic diagram for a logic circuit that implements the four bit integer square root function $y = \lfloor \sqrt{x} \rfloor$. 
In Figure \ref{root4bit_sidebyside} and in later figures and tables, $x_n$ is the $n^{th}$ binary digit of $x$ and $y_n$ is the $n^{th}$ binary digit of $y$.
The truth table for the circuit is shown in Table \ref{sqrttruthtable}. 

The right-hand side of Figure \ref{root4bit_sidebyside} shows how the structures in Figures \ref{tshapedregion_valves}, \ref{region2_valves} and \ref{region3_valves} can be used to implement this circuit. The logic gate symbols and excitable media channels in Figures \ref{root4bit_sidebyside} are shaded to show the correspondence between the two, and the key refers to the previous Figures in which the excitable media structures that are used here were introduced.

\begin{table}
\caption{Truth table for a 4-bit integer square root circuit.\label{sqrttruthtable}}
\begin{ruledtabular}
\begin{tabular}{|c|c|c|c|c|c|c|c|}
\hline
$x$ & $x_3$ & $x_2$ & $x_1$ & $x_0$ & $y = \lfloor \sqrt{x} \rfloor$ & $y_1$ & $y_0$ \\
\hline
$0$ & $0$ & $0$ & $0$ & $0$ & $0$ & $0$ & $0$ \\
$1$ & $0$ & $0$ & $0$ & $1$ & $1$ & $0$ & $1$ \\
$2$ & $0$ & $0$ & $1$ & $0$ & $1$ & $0$ & $1$ \\
$3$ & $0$ & $0$ & $1$ & $1$ & $1$ & $0$ & $1$ \\
$4$ & $0$ & $1$ & $0$ & $0$ & $2$ & $1$ & $0$ \\
$5$ & $0$ & $1$ & $0$ & $1$ & $2$ & $1$ & $0$ \\
$6$ & $0$ & $1$ & $1$ & $0$ & $2$ & $1$ & $0$ \\
$7$ & $0$ & $1$ & $1$ & $1$ & $2$ & $1$ & $0$ \\
$8$ & $1$ & $0$ & $0$ & $0$ & $2$ & $1$ & $0$ \\
$9$ & $1$ & $0$ & $0$ & $1$ & $3$ & $1$ & $1$ \\
$10$ & $1$ & $0$ & $1$ & $0$ & $3$ & $1$ & $1$ \\
$11$ & $1$ & $0$ & $1$ & $1$ & $3$ & $1$ & $1$ \\
$12$ & $1$ & $1$ & $0$ & $0$ & $3$ & $1$ & $1$ \\
$13$ & $1$ & $1$ & $0$ & $1$ & $3$ & $1$ & $1$ \\
$14$ & $1$ & $1$ & $1$ & $0$ & $3$ & $1$ & $1$ \\
$15$ & $1$ & $1$ & $1$ & $1$ & $3$ & $1$ & $1$ \\
\hline
\end{tabular}
\end{ruledtabular}
\end{table}

\subsection{Simulation}

A two-variable Oregonator equation \cite{field1974} adapted for light sensitivity and with a diffusion term
was used as an approximate model for the Belousov-Zhabotinsky reaction with applied illumination
on a planar surface in which reactants and products diffuse \cite{beato2003}:

\begin{minipage}{0.6\linewidth}
\[
\frac{\partial u}{\partial t} = \frac{1}{\epsilon} (u-u^2-(fv+\phi(r,t)) \frac{u-q}{u+q}) + D_u \nabla^2 u
\]
\end{minipage}
\hspace{0.1\linewidth}
\begin{minipage}{0.3\linewidth}
\[
\frac{\partial v}{\partial t} = u - v
\]
\end{minipage}

\vspace{5mm}

The variables $u$ and $v$ represent the local concentrations of activator and inhibitor respectively.
Parameter $\epsilon$ sets up a ratio of time scale for the variables $u$ and $v$, $q$ is a 
scaling parameter dependent on the rates of activation/propagation and inhibition, $f$ is 
a stoichiometric factor. $\phi(r,t)$ is the rate of inhibitor production, which is proportional to the level
of illumination. This varies over space $r$, and for valve locations $\phi(r,t)$ also varies with time $t$.

The system was integrated using the Euler method with a five-node Laplace operator for the diffusion term. The time step $\Delta t$ was $0.001$ and the
grid point spacing $\Delta x$ was $0.1$ (which corresponds to 0.1mm in the experimental setup). The following values were chosen so as to give behaviour qualitatively similar to that observed in experiments: $\epsilon = 0.033$, $f = 1.4$, $q = 0.002$, $D_u = 0.067$. For locations and times at which $\phi(r,t)$ is excitable, $\phi(r,t) = 0.055$. For other locations, $\phi(r,t) = 0.1$. This system provides a good enough simulation of the experimental setup to help establish what
types of behaviour are likely to be achievable experimentally.

\begin{figure}
\includegraphics[width=2.0in]{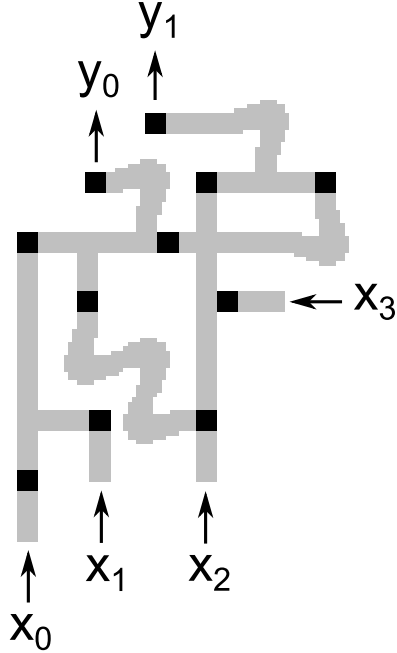}
\caption{Pattern for the 4-bit square root circuit. Grey areas are areas of constant excitability. Dark areas 
are valves which alternate between excitable and unexcitable. White areas are unexcitable.\label{sqrtmask}}
\end{figure}

The pattern for the 4-bit square root circuit is shown in Figure \ref{sqrtmask}. Each channel in Figure \ref{sqrtmask} is 15 grid points.
For grey areas of the pattern (excitable channels) $\phi(r,t) = 0.055$, for white areas (unexcitable regions) $\phi(r,t) = 0.1$, and for black
areas (valves) $\phi(r,t)$ is a square wave with period $P$ and a 50\% duty cycle:
\[
\phi(r,t) = 0.055 \textrm{ when } nP < t \le nP + P/2 \textrm{ for some integer n, otherwise } \phi(r,t) = 0.1
\]
By measuring the number simulation steps it took for a wave to travel from  the valve nearest $x_0$ to the next valve directly in line with $x_0$, $P$ was set to $15.5$ (i.e. 15500 simulation steps). Corners in the pattern mask are shaped so as to make the time for wave propagation along two channel segments at
right angles to each other the same as when the two channel segments are placed together in a straight line.

Figure \ref{sqrtexpsim} at the end of section \ref{experimental} shows snapshots from the simulation alongside experimental results for the cases $x=3$, $x=6$ and $x=15$. A video showing a simulation of all 15 cases can be found at the URL given in reference \cite{simulation-video}.

\subsection{Experimental implementation \label{experimental}}

\subsubsection{Experimental setup}

Sodium bromate, sodium bromide, malonic acid, sulphuric acid and tris(bipyridyl) ruthenium(II)
chloride were purchased (Sigma-Aldrich, U.K., BH12 4QH) and used as received unless
stated otherwise.

The catalyst free reaction mixture was freshly prepared in a 30ml continuously-fed stirred tank
reactor (CSTR), which involved the in-situ synthesis of stoichiometric bromomalonic acid
from malonic acid and bromine generated from the partial reduction of sodium bromate. This CSTR
in turn continuously fed a thermostated open reactor with fresh catalyst-free BZ reaction
mixture in order to maintain a non-equilibrium state. The final composition of the
catalyst free reaction solution in the reactor was 0.42M sodium bromate, 0.19M malonic acid,
0.64M sulphuric acid, and 0.11M sodium bromide. The residence time was 30 minutes. The open reactor was surrounded by a water jacket thermostated at 25$^\circ$C.

\begin{figure}
\includegraphics[width=4.0in]{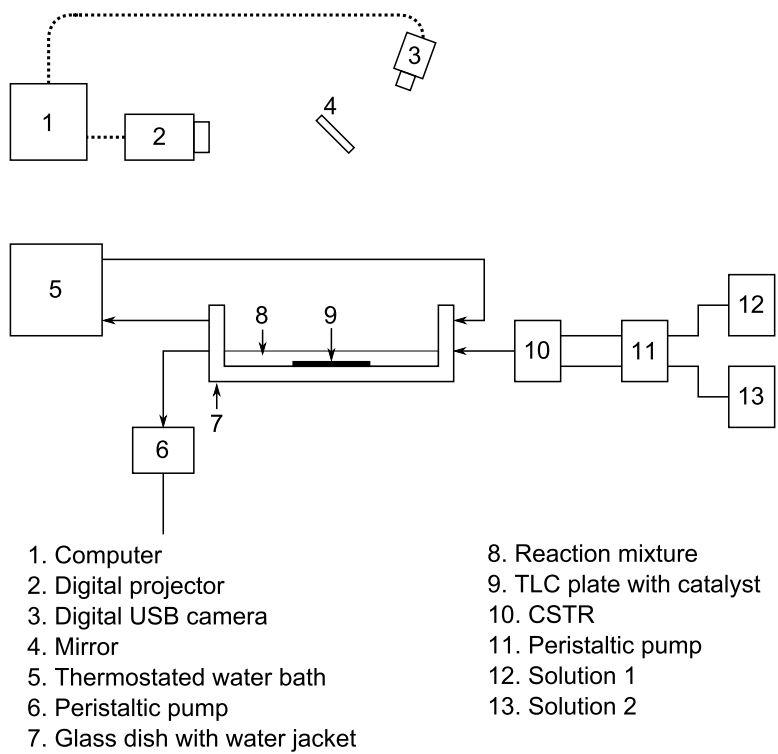}
\caption{A block diagram of the experimental setup.\label{experimental_setup}}
\end{figure}

Peristaltic pumps (Model Sci-Q 400, Watson Marlow Ltd. UK. TR11 4RU) were used to pump
the reaction solution into the reactor and remove the effluent. A diagrammatic
representation of the experimental setup is shown in Figure \ref{experimental_setup}. Solution 1 is 0.84M sodium bromate, 0.38M malonic acid. Solution 2 is 1.28M sulphuric acid, 0.22M sodium bromide. 

 The ruthenium catalyst solution was prepared by recrystallising \ce{Ru(bpy)3SO4} from the
chloride salt with sulphuric acid and then dissolving 0.5g of crystal \ce{Ru(bpy)3^{3+}} in 18ml of deionised water to make 0.025M \ce{Ru(bpy)3^{3+}} solution.

A thin layer chromatography (TLC) plate (T-7270, Sigma-Aldrich, UK) was used as a substrate for the ruthenium catalyst.
The silica gel layer was 250\si{\micro\meter} thick, the mean particle size was 2-25\si{\micro\meter} and the mean pore diameter was 60\si{\angstrom}.  As supplied, the plate measured 5cm by 20cm and was
divided (by scoring and then snapping over the side of a bench) into sections measuring 5cm by 6.5cm to fit into a Petri dish. A TLC plate section was placed into a Petri dish containing a mixture of 14\si{\milli\liter} de-ionised
water and 0.9\si{\milli\liter} ruthenium catalyst solution. The plate was left in the Petri dish overnight (for approximately 18 hours) and then removed and placed into
a Petri dish containing de-ionised water for 30 minutes to rinse off excess catalyst. The plate was then moved to a third Petri dish containing de-ionised water for a further 30 minutes before being used.
During the course of a 5-hour experimental session the catalyst slowly washed out of the TLC plate and as a consequence the plate became less excitable as the experiment progressed. 

The digital projector was a Sanyo PLC-XT21L. The brightness setting of the projector was gradually reduced during the course of a 5-hour experimental session to compensate for the washing out of the catalyst, so as 
to make sure that waves were able to propagate in the fully excitable regime in all dark areas in the projected image, but unable to propagate in light areas.

During the attempts to achieve a uniform subexcitable regime across the whole plate discussed in section \ref{interacting_waves}, it was found that the level of excitability varied across the plate. It is believed that these variations were caused by differences in the amount of catalyst in different parts of the plate. However, whereas the range of conditions needed to obtain a subexcitable regime is very narrow, the range needed to obtain a consistent regime in which the substrate is fully excitable in dark areas but unexcitable in light areas is much broader, and the behaviour of propagating waves in this regime is not very sensitive to small changes in illumination or excitability.

The light intensity in dark areas of the projected image is estimated at $< 0.1$\si{\milli\watt\per\square\centi\meter}, and the light
intensity in light areas ranged from about 7\si{\milli\watt\per\square\centi\meter} at the beginning of an experimental run to about 4\si{\milli\watt\per\square\centi\meter} at the
end of an experimental run. These estimates are based on measurements with a Centronix OSD1-5T photodiode,
using the spectral response curve in the manufacturers datasheet and assuming that the white light produced
by the digital projector has a peak wavelength at 550nm and has a power spectrum that is approximately symmetrically distributed around this peak in the range 425nm-750nm.

The projected circuit pattern mask was similar to that used during simulation (shown in Figure \ref{sqrtmask}), but scaled down by a factor of 3. The dimensions of the projected circuit pattern were 33.5mm by 36mm. The width of each channel in the projected image was 1.5mm. Each pixel in the projected image measured 0.3mm by 0.3mm.

The period to use for the valves in the projected image (corresponding to $P$ in the simulation) was determined at the beginning of the experimental
session by timing the passage of a wave from one valve to another. A period of 280 seconds was used, so every valve was open for 140 seconds, then closed for 140 seconds, and so on.

Images were captured once every 5 seconds with a Lumenera Infinity 2 USB camera fitted with a 455nm narrow bandpass interference filter.

\subsubsection{Results \label{results}}

\begin{figure}
  \subfloat[$x=3$ experiment, half-way through.]{\label{fig:3mid}\includegraphics[width=1.35in]{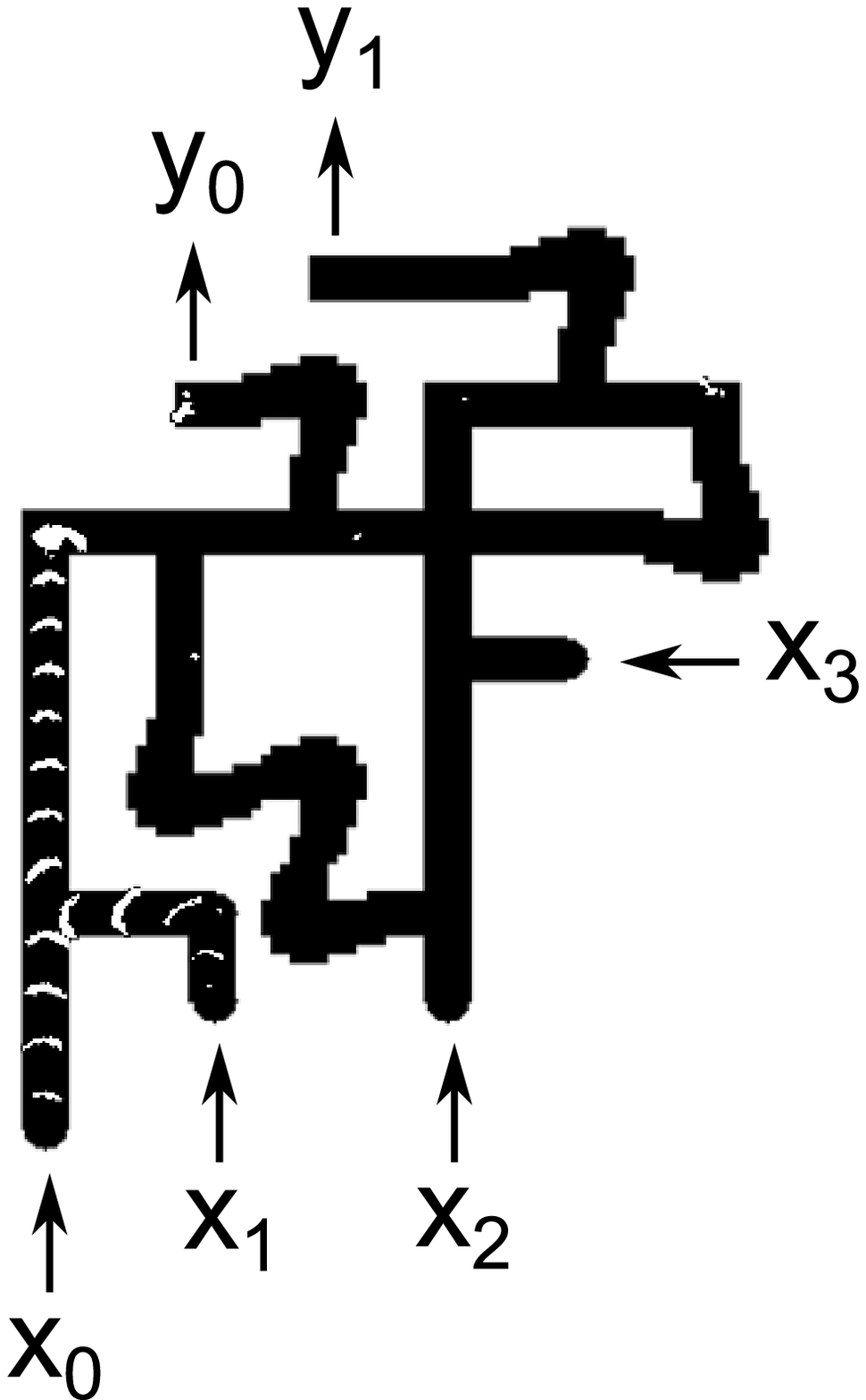}}   
  \subfloat[$x=3$ simulation, half-way through.]{\label{fig:3midsim}\includegraphics[width=1.35in]{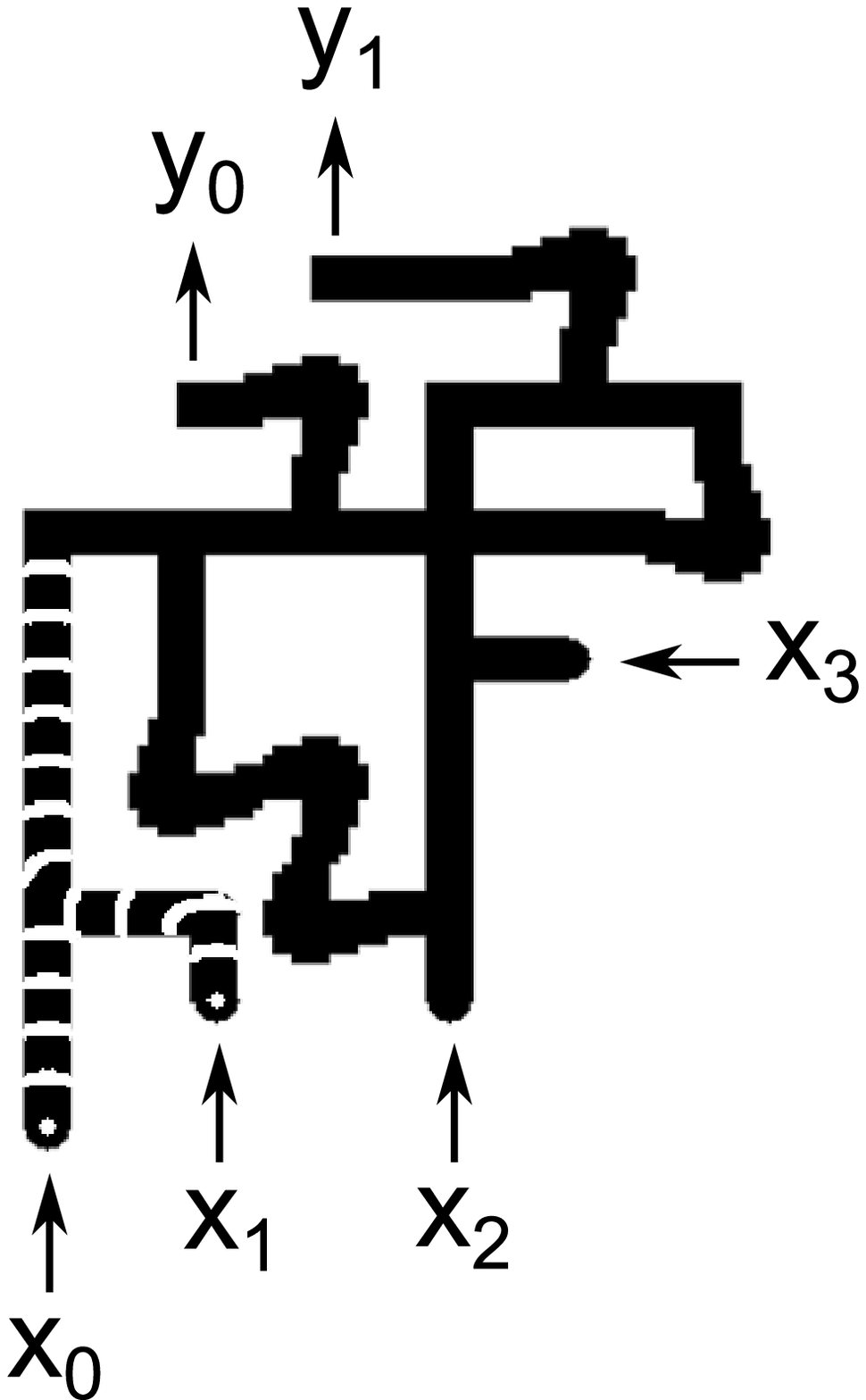}}   
  \hspace{6mm}
  \subfloat[$x=3$ experiment, complete.]{\label{fig:3all}\includegraphics[width=1.35in]{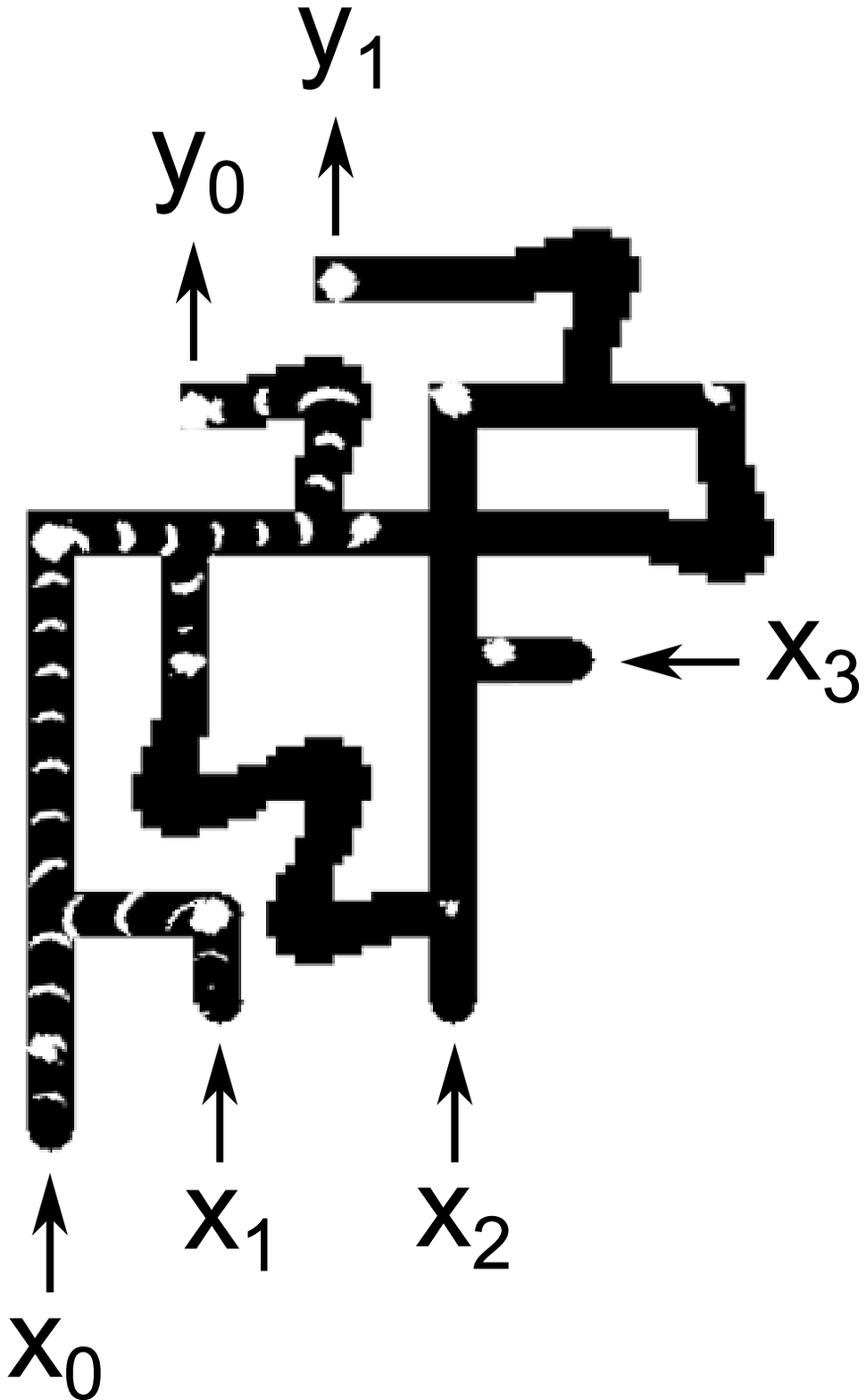}}   
  \subfloat[$x=3$ simulation, complete.]{\label{fig:3allsim}\includegraphics[width=1.35in]{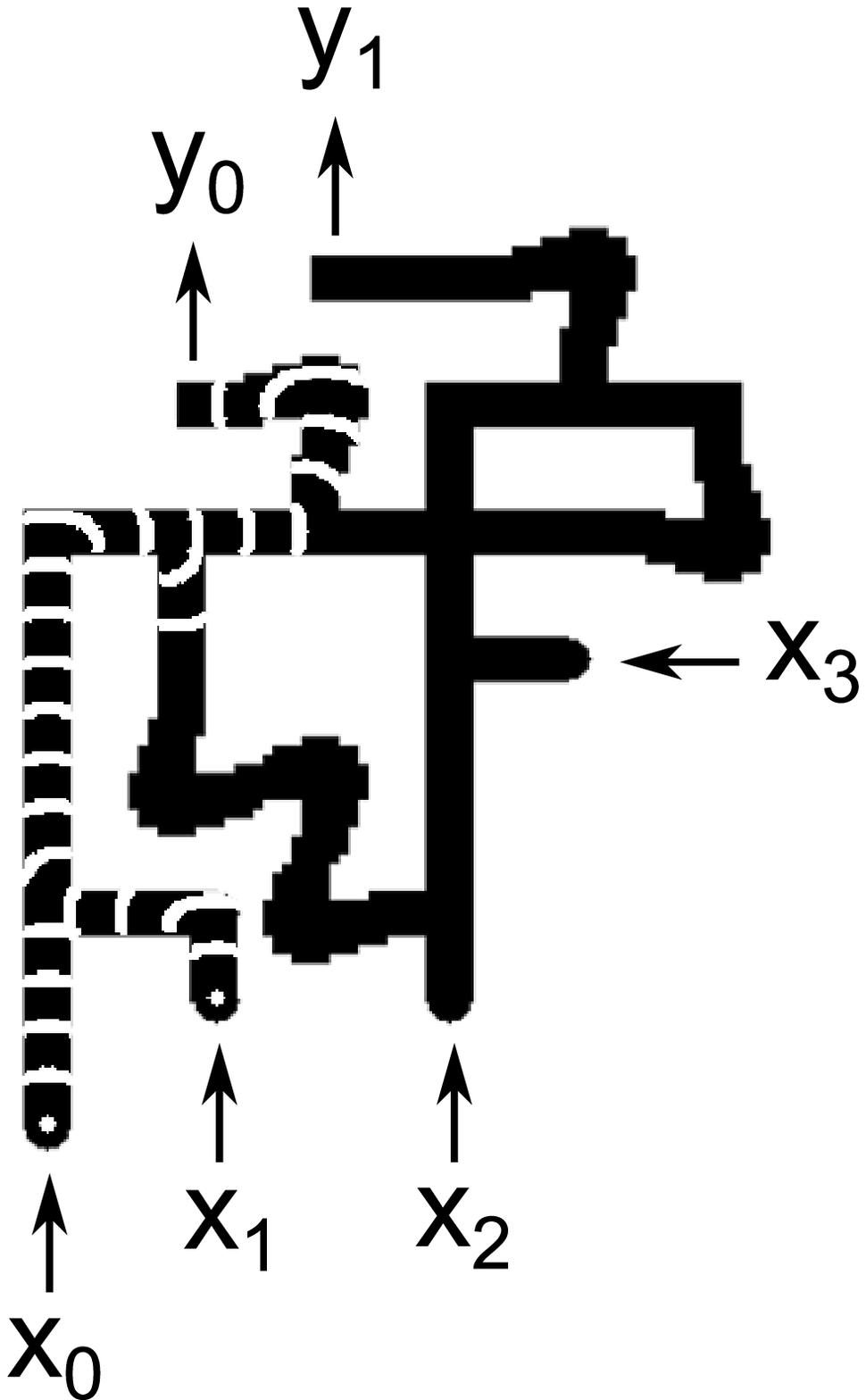}}   
  \hspace{6mm}
  \subfloat[$x=6$ experiment, half-way through.]{\label{fig:6mid}\includegraphics[width=1.35in]{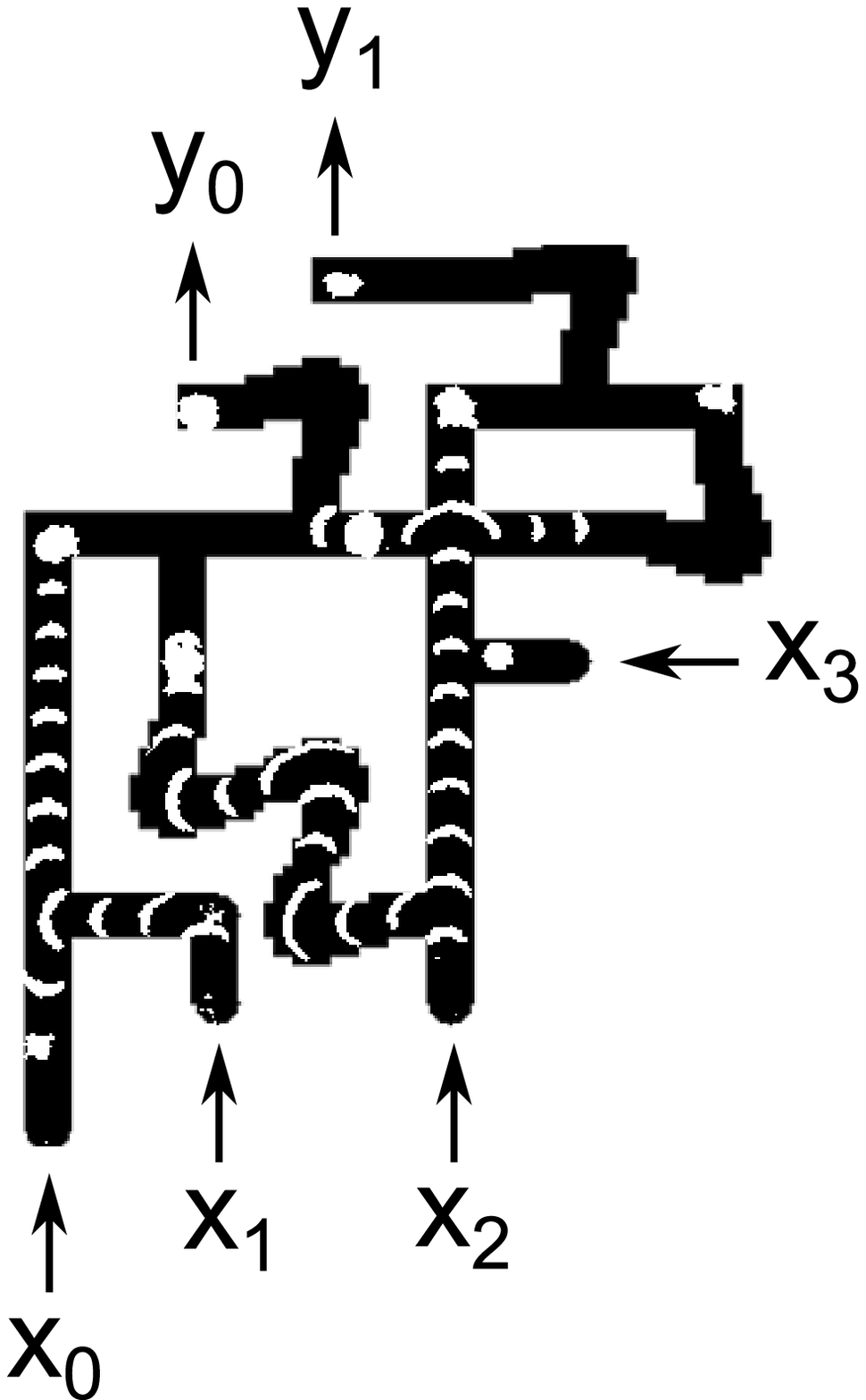}}   
  \subfloat[$x=6$ simulation, half-way through.]{\label{fig:6midsim}\includegraphics[width=1.35in]{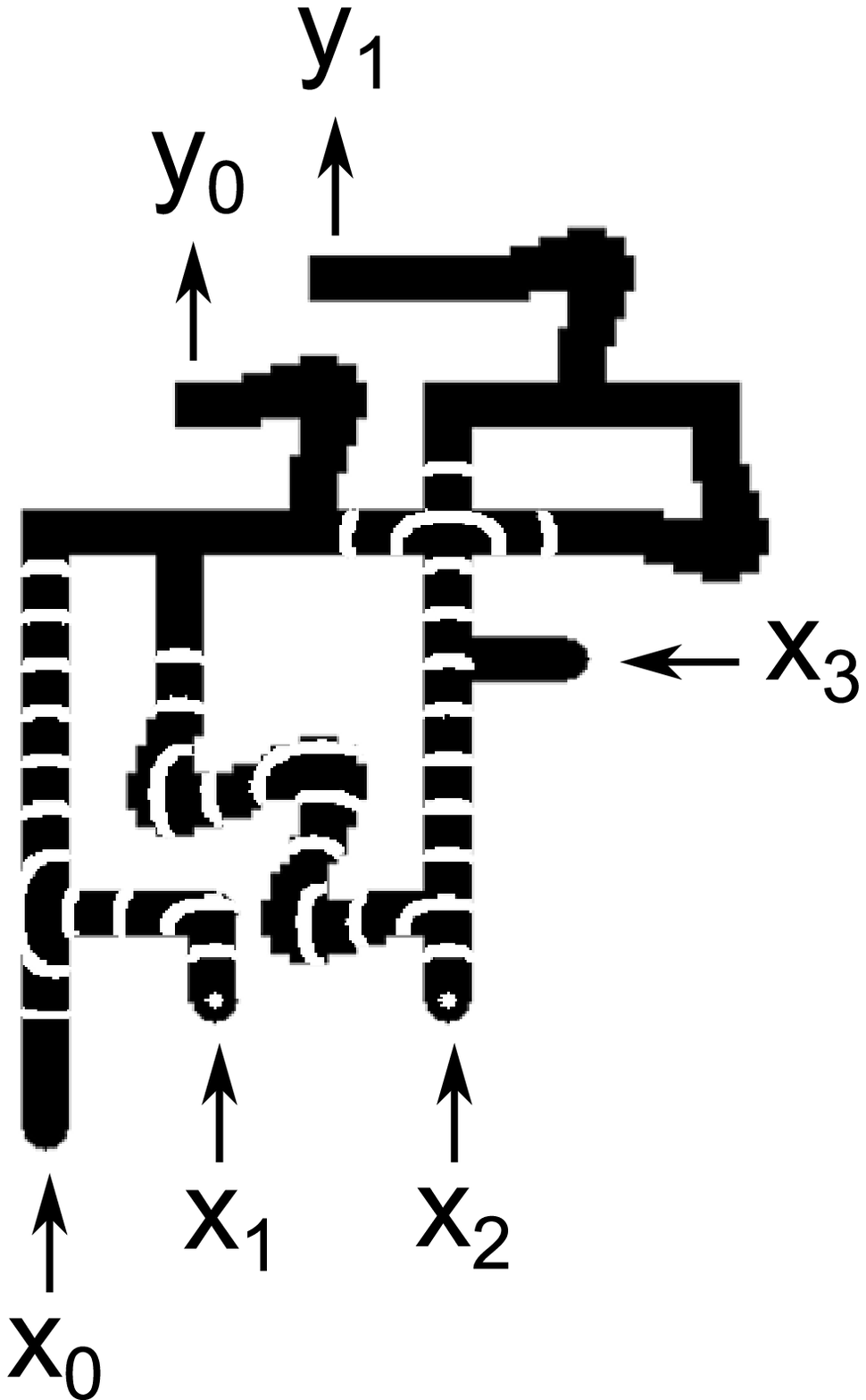}}   
  \hspace{6mm}
  \subfloat[$x=6$ experiment, complete.]{\label{fig:6all}\includegraphics[width=1.35in]{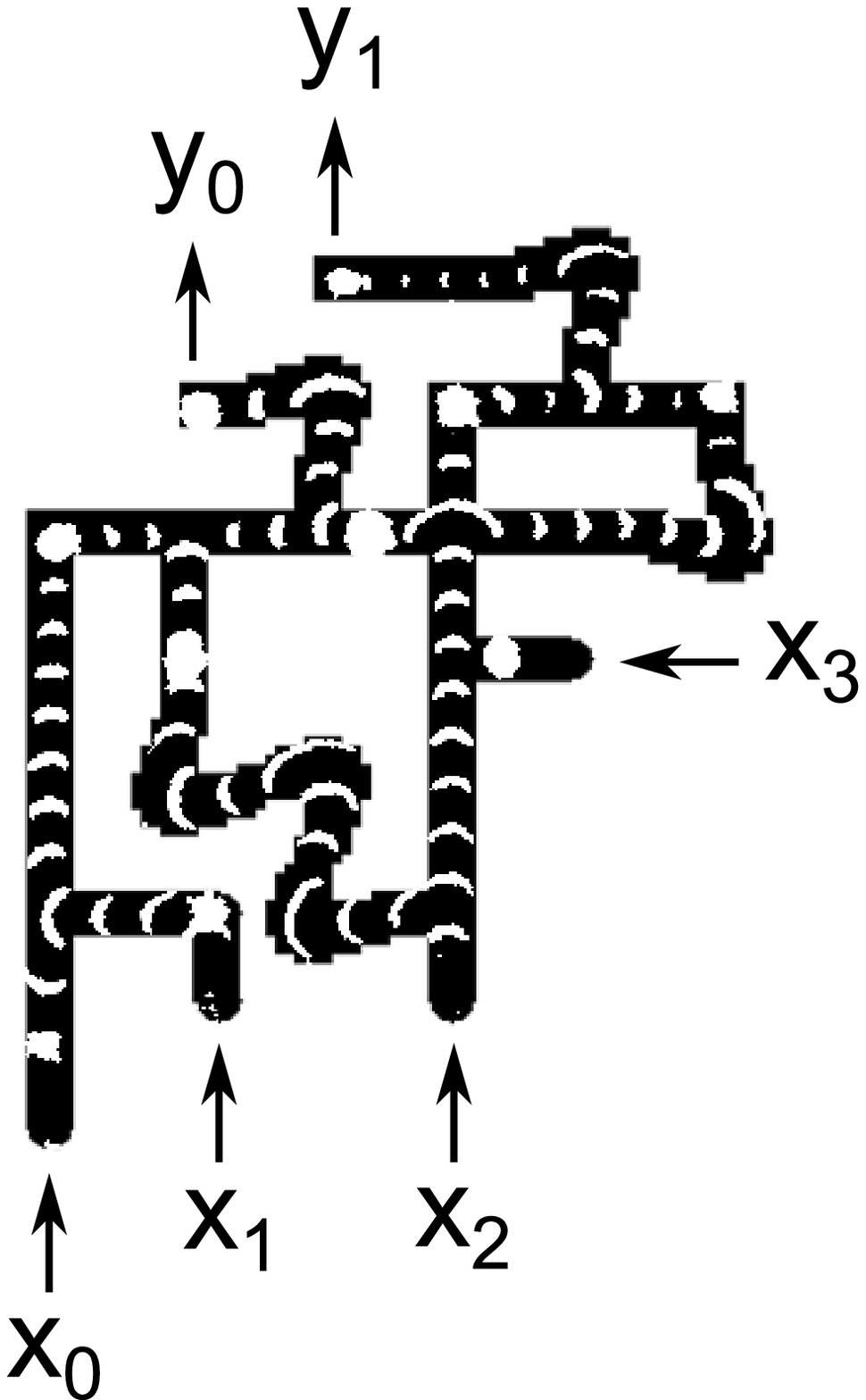}}   
  \subfloat[$x=6$ simulation, complete.]{\label{fig:6allsim}\includegraphics[width=1.35in]{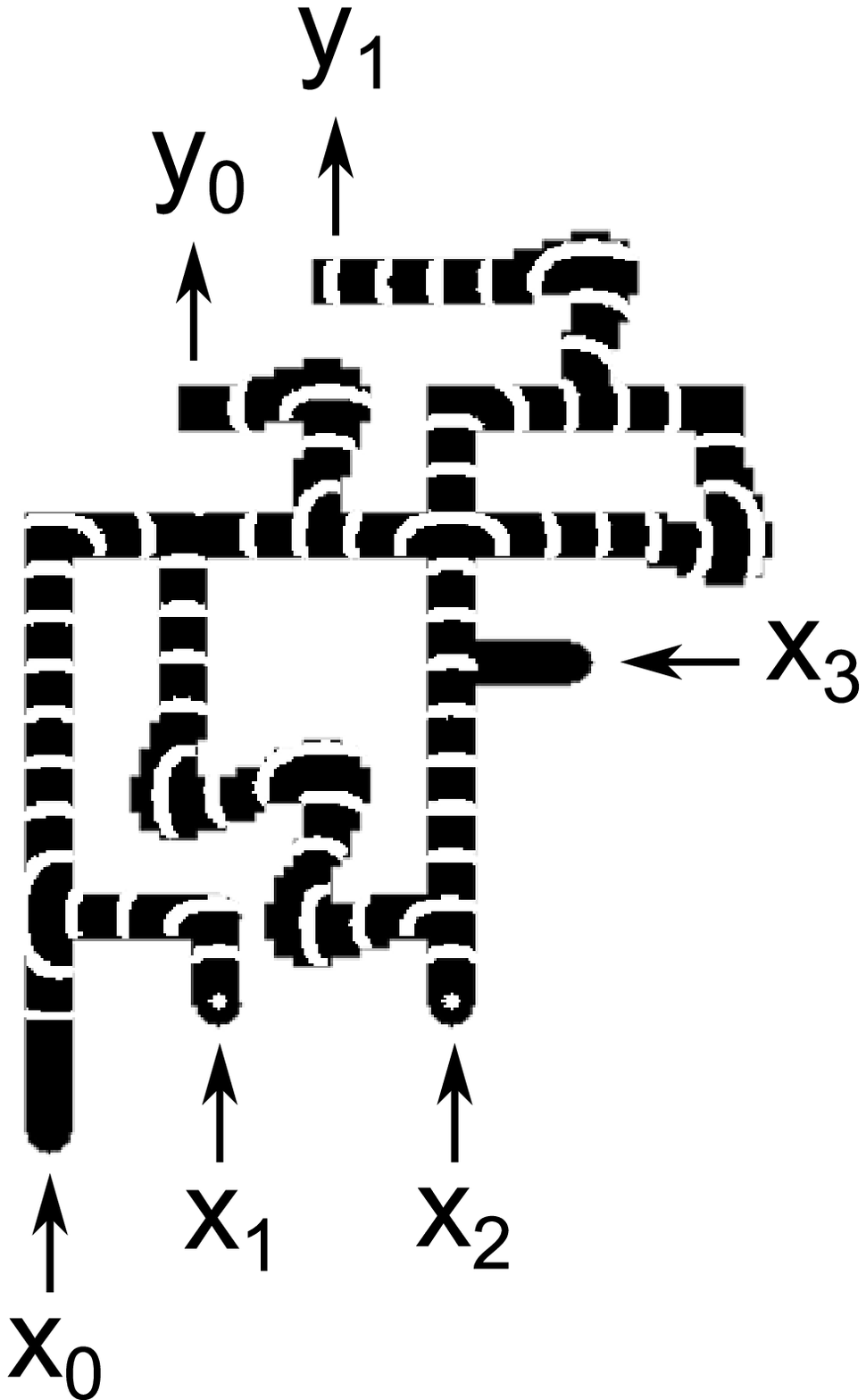}}   
  \hspace{6mm}
  \subfloat[$x=15$ experiment, half-way through.]{\label{fig:15mid}\includegraphics[width=1.35in]{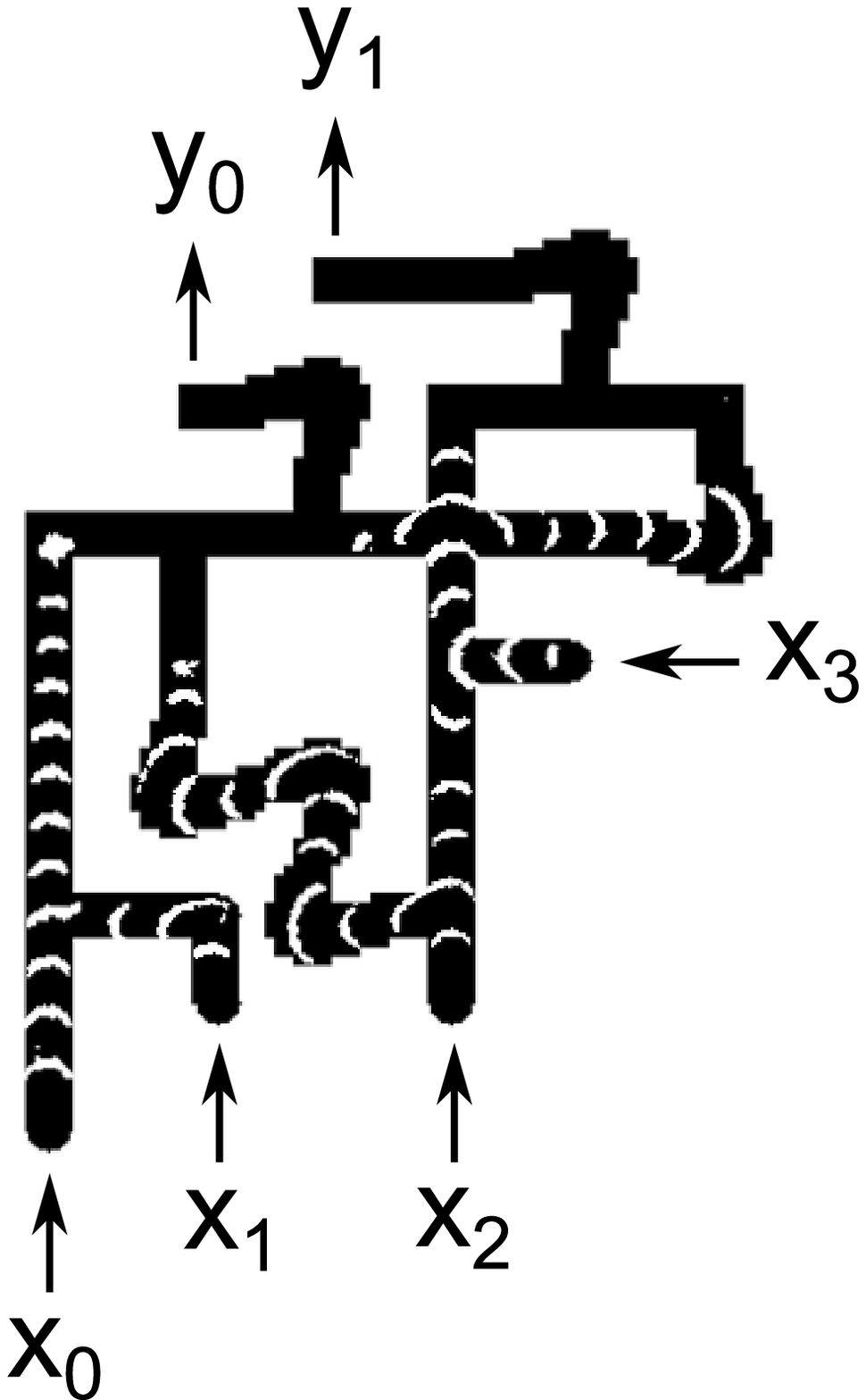}}   
  \subfloat[$x=15$ simulation, half-way through.]{\label{fig:15midsim}\includegraphics[width=1.35in]{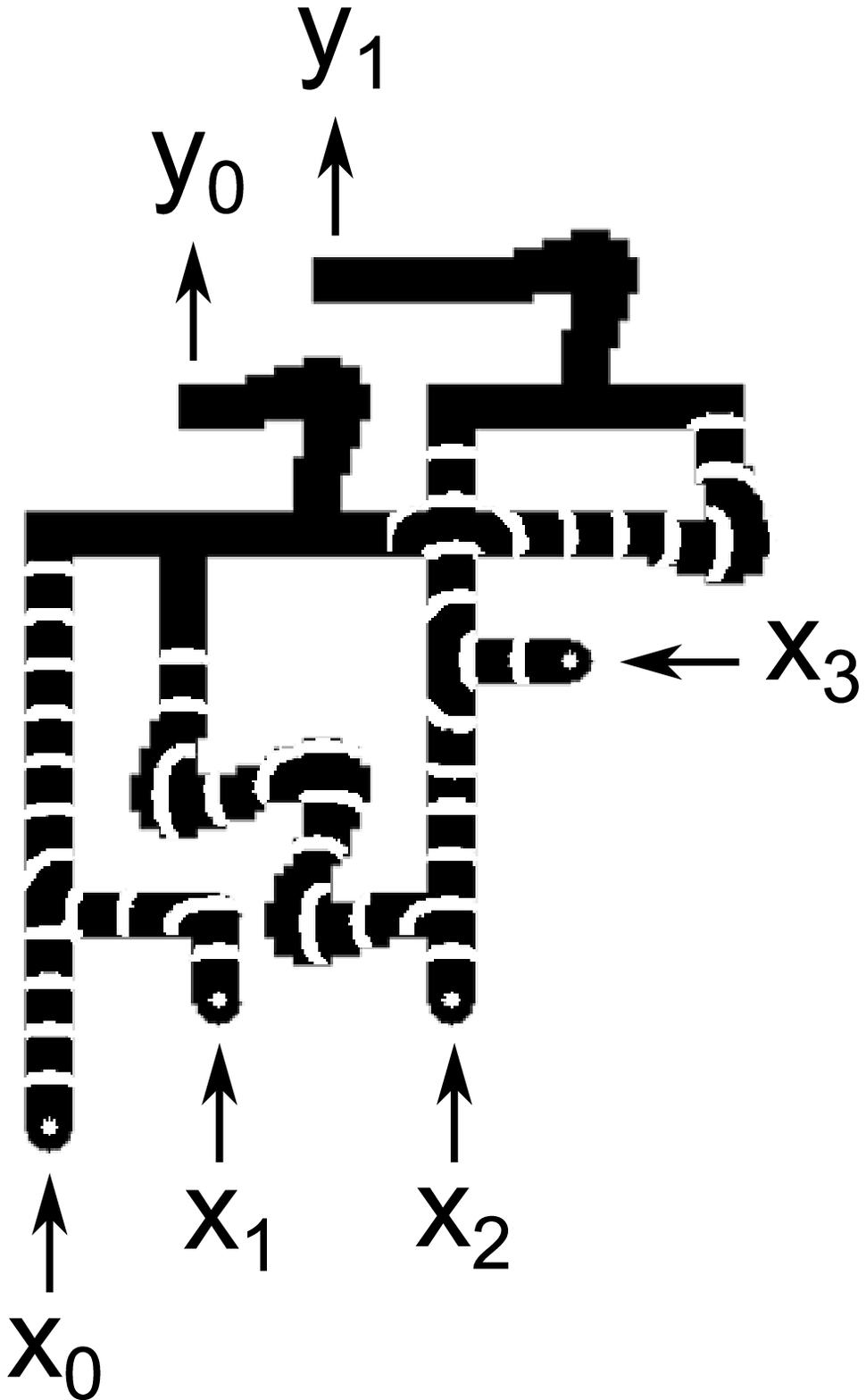}}   
  \hspace{6mm}
  \subfloat[$x=15$ experiment, complete.]{\label{fig:15all}\includegraphics[width=1.35in]{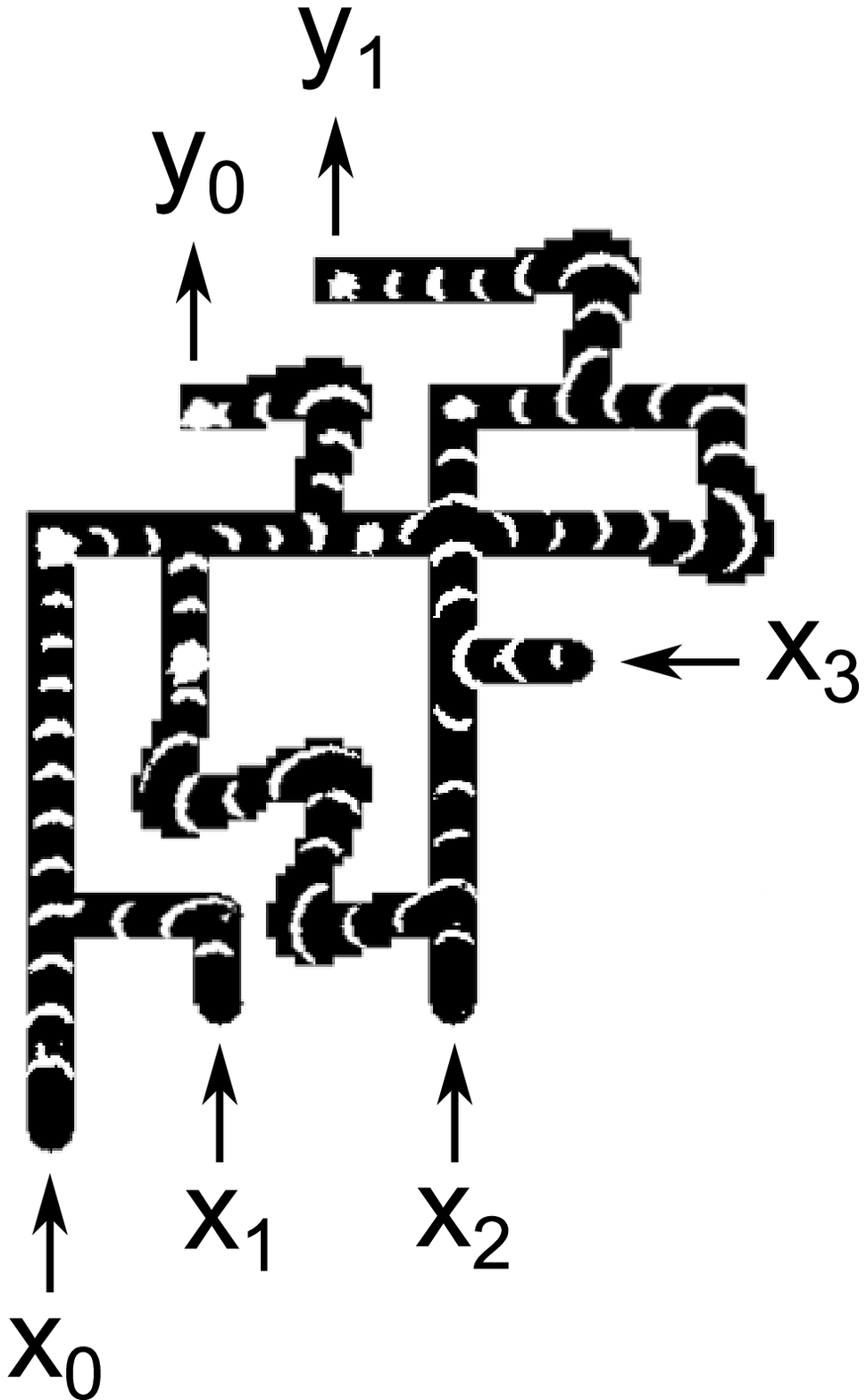}}   
  \subfloat[$x=15$ simulation, complete.]{\label{fig:15allsim}\includegraphics[width=1.35in]{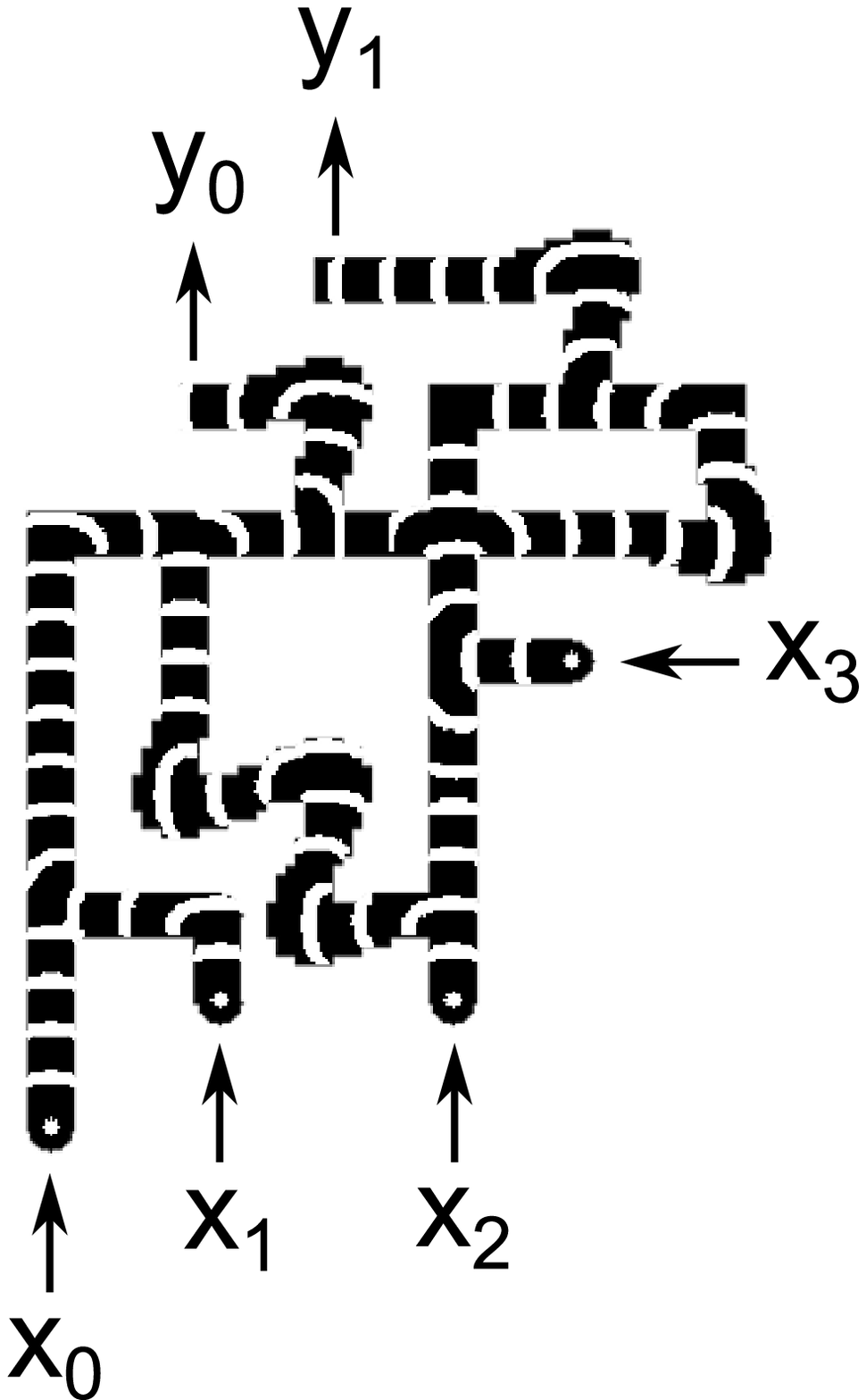}}   
  \caption{Experiment and simulation for cases $x=3,6$ and $15$. Note that for the $x=6$ case, no wave emerges from $y_0$ because the valve at $y_0$ is closed when a wave reaches it.\label{sqrtexpsim}}
\end{figure}

The propagation speed for waves was 6.7\si{\milli\meter\per\minute} and it took 10 minutes to for waves to propagate from inputs to outputs.
Figure \ref{sqrtexpsim} compares snapshots taken from experimental runs with the corresponding snapshots from simulations. The cases $x=3$, $x=6$ and $x=15$ are shown. For each case, one pair of images shows the circuit halfway through its operation, and another pair of images shows the circuit at the end of its operation. The experimental snapshots were produced by taking the difference of images 30 seconds apart.

Reference \cite{experiment-video} contains a URL for a video showing 7 cases ($x=3,x=4,x=6,x=8,x=9,x=12,x=15$) which demonstrate the circuit functioning correctly. Because inputs $x_0$ and $x_1$ are ORed together immediately, these 7 cases fully exercise every gate in the circuit.

The main problems encountered during experiments were:
\begin{itemize}
\item It was not easy to arrange for perfectly synchronised waves to enter the inputs. However, because each valve is open from $0.25P$ before the time that a wave is expected to pass through it until $0.25P$ after this time, input waves can pass through the inputs up to 70 seconds before or after they are expected (corresponding to a distance of $\pm{}7.8$mm) before this affects the output of the circuit.

\item The wave speed was not identical across the whole plate. In a more complex circuit with many
gates cascaded together,
small differences in timing will accumulate. Eventually this will lead to some waves being blocked when they should not be, or waves passing through a valve when they should be blocked. Assuming that waves travel 1 unit of length in 1 unit of time, that valves are open at $t_0+2n - 0.5 < t \le t_0+2n + 0.5$ for any integer $n$ and closed at other times, that all input-output paths for a complex circuit are the same length $L$, and that all input waves enter the circuit at $t_0$, then the average speed of a wave must be between $\frac{L}{L+0.5}$ and $\frac{L}{L-0.5}$ to avoid any waves being blocked by valves when they should not be, and vice-versa. No matter how good the experimental
conditions are, for a large enough $L$ this requirement will not be met. It may be possible to overcome
this problem by periodically resynchronising waves to a time reference. One way of doing this is described in section \ref{resync}.
\end{itemize}

\section{Discussion and Conclusion\label{discussion}}

We have designed a 4-bit input, 2-bit output integer square root circuit in BZ excitable media and successfully implemented the circuit experimentally. This is the most complex circuit that has been implemented in BZ excitable media to date.
We believe that one of the reasons why the implementation of this circuit was straightforward is that it only makes use of two of the many phenomena that BZ excitable media can exhibit: constant speed wave propagation and the annihilation of colliding wavefronts. Neither of these phenomena are particularly sensitive to variations in experimental conditions. For the same reason, there is also a reasonably good correspondence between qualitative simulations and experiments.

From a theoretical point of view, it is interesting and useful to consider all possible phenomena in excitable media that can be used for information processing. From a practical point of view, it is easier to implement those
systems that depend on the fewest phenomena: arranging experimental conditions so that all of the phenomena required by a system
are present simultaneously is easier if only a few phenomena are required than if many are required.
This was born out in our attempts to use the subexcitable regime for implementing fork structures, before
switching to the time-dependent wave selection scheme. 

Whereas previous work on logic circuits in structured excitable media has used static circuit patterns, we exploit the fact 
that the circuit pattern is projected using a digital projector to create periodic inhibiting valves which have
the effect of selecting waves based on the time that it takes them to propagate from one valve to another.
Because the behaviour of a gate in this scheme is dependent only on the distances between terminals, rather than on the precise geometry of the gate, paths making up a gate can often be folded up to make a compact circuit.
An interesting avenue for further research is to explore other ways in which patterns periodically repeating in time can be used  in excitable media information processing schemes.

\subsection{Auxiliary 1 inputs}

The most obvious and straightforward way to represent a Boolean value using events is to use the occurrence of an event within a defined
period of time to represent a Boolean 1 value, and the non-occurrence of an event within a defined period
of time to represent a Boolean 0 value. This is the representation which has been used by previous implementations of logic circuits in BZ excitable media, and is also used in this paper. One drawback of this method of representing Boolean values is that auxiliary 1 inputs are needed in order to perform certain functions, such as a NOT operation (a NOT gate can be replaced by an AND-NOT gate with a constant auxiliary 1 input fed to the non-inverting input). Below we consider other possible mappings between events and Boolean values that do not have this problem.

\subsection{Dual-rail logic}

One alternative representation is dual-rail logic, in which a Boolean value is represented
using a pair of channels. A Boolean 1 value is represented by the presence of a wave on one channel, and a Boolean 0 value is represented by the presence of a wave on the other channel \cite{vonneumann1956}.
An obvious disadvantage of a dual-rail representation is that it requires two channels rather than one for
each Boolean value. Generally, this disadvantage can be offset against the following potential advantages: 

\begin{itemize}
\item Inversion can be implemented by crossing over two signals. No auxiliary inputs are required.
\item Dual-rail logic can facilitate self-timing -- a Boolean value is always represented
 by one event or another: there is no need for an external time
  reference.
\item A set of complete Boolean functions with a dual-rail representation can be built from an
  incomplete set of single-rail Boolean functions (for example, one that does not have a NOT operation, or 
  which does not have a complete OR operation) \cite{stevens2011b}.
\end{itemize}

However, in the context of the work in this paper these do not seem to be significant advantages: the natural AND-NOT behaviour of a collision between waves means that inversion is already a straightforward operation (provided that an auxiliary Boolean 1 value is available), and self-timing cannot be fully realised without a mechanism that has a permanent internal state, which is difficult to implement efficiently in BZ excitable media. Since OR and AND-NOT gates (or just AND-NOT alone) already form a logically complete set of Boolean functions, the ability to construct a complete set of Boolean functions from an incomplete set is no advantage. 

\subsection{Using phase to represent Boolean values}

Another alternative is to use the phase of a wave to represent Boolean values. By \emph{phase} we mean the delay between a reference time $t_0$ and the occurrence of a wave in a particular location.  Consider a location $p$ on a channel. The propagation of a wave through $p$ at some time $t_0$ (which we will call a leading wave) represents
a logic 1, and the propagation of a wave through $p$ at time $t_0+1$ (which we will call a lagging wave) represents a logic 0.

This representation makes the construction of an OR gate very straightforward. A T-shape without any valves can be used as an OR gate because with this representation there is no way that a signal can propagate backward through an input: if 
both input waves are either leading or lagging then they will merge, and the output will follow the inputs, delayed by 2 time units. Otherwise one input wave will arrive at $t_0$ and will begin propagating backward through the other input (as well as propagating to the output). But it will encounter the other input wave, arriving at $t_0+1$, and will annihilate with it.

\begin{figure}
\includegraphics[width=1.5in]{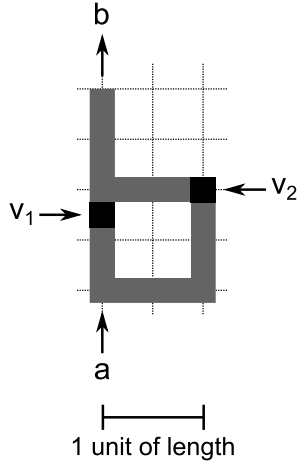}
\caption{A NOT gate which turns a leading wave into a lagging wave, and a lagging wave into a leading wave.\label{phase_inverter}}
\end{figure}

A NOT gate in this representation can be constructed as in Figure \ref{phase_inverter}. For the case when a wave enters the NOT gate at $a$ at $t_0$ (when both $v_1$ and $v_2$ are open), the wave will reach $v_1$ at $t_0+1$ when 
$v_1$ is closed, and will not pass through. But it will reach $v_2$ at $t_0+2$ when $v_2$ is open, and will
end up exiting the inverter at $b$ at $t_0+4$. 
For the case when a wave enters the NOT gate at $a$ at $t_0+1$ (when both $v_1$ and $v_2$ are closed), the
wave will reach $v_1$ at $t_0+2$ when $v_1$ is open, and will pass through to reach $b$ at $t_0+3$. The NOT gate turns a leading wave into a lagging wave and a lagging wave into a leading wave.
Between them, the NOT gate and the OR gate can be used to make any desired logic circuit.

\subsection{Crossover}

Although the 4-bit integer square root circuit described in this paper does not require any wires to be crossed over, this section describes a structure that may be used to implement crossing wires when this is required.
Dewdney \cite{dewdney1979} showed how to implement a planar crossover circuit using any functionally complete set of logic gates. A crossover circuit can be made from 10 AND-NOT gates. 

\begin{figure}
  \subfloat[A crossover structure for $x$ and $y$. $x$ will propagate to $x'$, alternatively $y$ will propagate to $y'$, but the crossover fails if both $x$ and $y$ occur at the same time.]{\label{fig:crossover1}\includegraphics[width=2.5in]{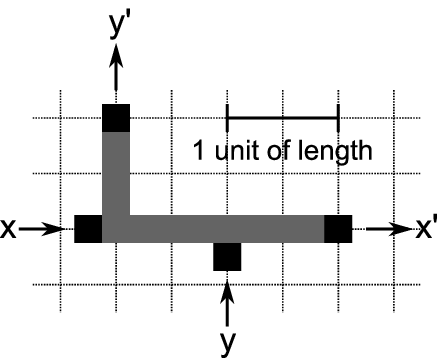}}   
  \hspace{6mm}
  \subfloat[A crossover structure for $x$ and $y$ which will work even if $x$ and $y$ occur at the same time.]{\label{fig:crossover2}\includegraphics[width=2.5in]{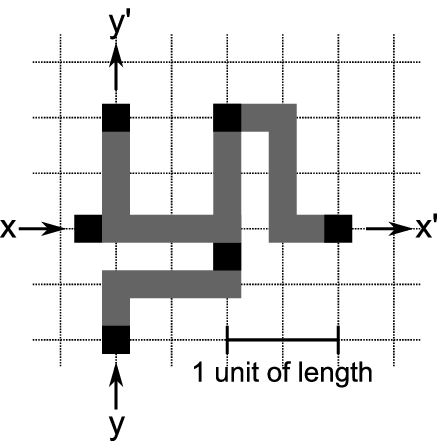}}   
  \caption{Structures for crossing over $x$ and $y$.}  
\end{figure}

We can do better than this however. Figure \ref{fig:crossover1} shows a crossover that will permit $x$ to propagate to $x'$ (but not to $y$ or $y'$), or alternatively permit $y$ to propagate to $y'$ (but not to $x$ or $x'$), so long as both $x$ and $y$ do not occur at the same time. This crossover structure works because the $x'$ output is the only terminal an even distance from $x$, and the $y'$ output is the only terminal an even distance from $y$. If we wish to deal with the case when $x$ and $y$ may both occur at the same time, we can modify this structure by delaying any wave input at $y$ by two time units. To keep the output signals synchronised, any
wave output at $x'$ is also delayed by two time units. This is shown in Figure \ref{fig:crossover2}.

\subsection{A general abstract description of time-dependent wave selection}

The reason why we are able to make such a simple crossover mechanism -- much simpler than one made from 10 AND-NOT gates -- is that we are working at a level of abstraction lower than that of Boolean logic gates. Time-dependent wave selection can be used to implement logic gates, but it can also be used to implement other structures more directly than via logic gates.

A general abstract description of the behaviour of waves propagating between valves in a region of BZ excitable medium is stated below. The internal distance $| \vec{pq} |$ between any two points $p$ and $q$ within a region is 
the length of the shortest path between $p$ and $q$ that lies within the region. Here we assume that the internal distance between any pair of terminals is an integral length and that the region contains no unexcitable holes.

\begin{figure}
\includegraphics[width=3in]{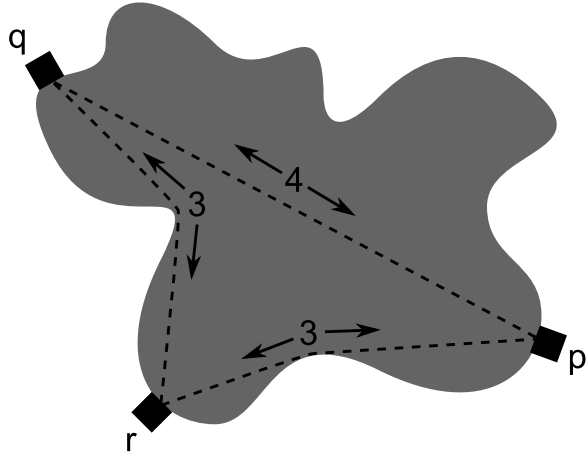}
\caption{A an arbitrarily shaped region of excitable media with three terminals with valves. Distances between pairs of valves are shown.\label{anyshaperegion}}
\end{figure}

A region may be of any shape, and may contain any number of terminals on its boundary, with each terminal
having a valve that is only open at even times.  A wave starting from a given terminal $p$ at
an even time $t$ will propagate to a terminal $q$ at time $t + | \vec{pq} |$ so long as the distance $| \vec{pq} |$ is even,
and there is no wave that begins 
from any terminal $r$ for which $| \vec{rq} | < | \vec{pq} |$ at an even time greater than or equal to $t - | \vec{rp} |$ but less than
$t + | \vec{pq} | - | \vec{rq} |$ (no waves can occur at $r$ at odd times because the valve at $r$ will block them).
This is illustrated in Figure \ref{anyshaperegion}. In Figure \ref{anyshaperegion} a wave from $p$ at $t_0$ will propagate
to $q$ at $t_0+4$ unless there is a wave from $r$ at $t_0-2$ or $t_0$.

To simplify the process of designing structures, in this paper we have only used regions with edge-lengths that are
multiples of half a unit length, and any pair of edges is either parallel or orthogonal. To reduce the
space that structures occupy, we have used thin regions. 

\subsection{Resynchronisation \label{resync}}

\begin{figure}
\includegraphics[width=2.5in]{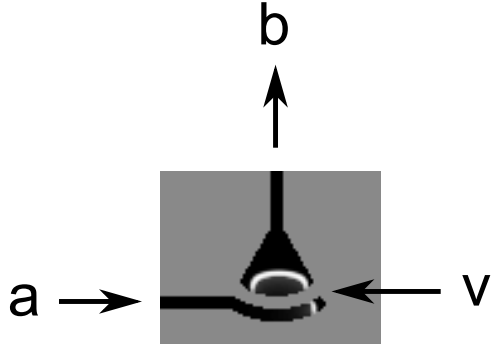}
\caption{A mechanism that will synchronise waves with valve timing. All points along valve $v$ are the same distance from $b$, so that a wave originating from $a$ will reach $b$ at a constant time after $v$ opens, regardless of the exact time at which it entered $a$.\label{synchroniser}}
\end{figure}

In section \ref{results} we discussed the need to resynchronise waves to a time reference to remove
small variations in wave propagation time that may accumulate for large logic circuits. Figure \ref{synchroniser} shows one possible mechanism for doing this. This mechanism uses the time of valve opening as a reference time. Waves exiting the mechanism are always at a known offset from the valve opening time. The entire curved region labelled $v$, between the two dark regions, is a valve in the shape of a circular arc.  When valve $v$ is closed, a wave entering $a$ will propagate along the curved channel beneath $v$. When the valve opens, the wave will propagate through the valve and head towards $b$. Because all points along valve $v$ are equidistant from $b$, the wave will reach $b$ at a constant time after $v$ opens, regardless of the exact time at which it entered $a$. 

\subsection{Parallels with neural information processing}

The phenomena used to implement the scheme described here have some similarities to phenomena that are
present in neural circuits. Both involve the propagation of excitations at a predictable speed, and both  have regions in which excitations can be inhibited. Among the large number of things that we do not yet understand about neural information processing is the range of roles that inhibitory synapses play. The scheme described here suggests a new possibility: it may be possible to use neurons to construct circuits in which thresholding plays no role at all in information processing. In such a scheme neurons serve simply
as repeaters, and as the providers of a medium in which excitations can propagate and interact. Periodically spiking inhibitory neurons with axons leading to inhibitory synapses on other neurons could play the same role that valves play in this paper.

During the evolution of nervous systems, we do not know whether the physiology of neurons and the phenomena that are exhibited by neurons were selected for because they are well suited to their role as components of evolvable, adaptable information processing systems, or whether their evolution was highly constrained by other factors. If we assume the former, we should consider systematically evaluating models which incorporate one or more of the phenomena exhibited by neurons for their potential use in evolutionary computing schemes (as has been done in \cite{stiefel2007,bull2009}). The work in this paper is one such model.

\section{Acknowledgements}

William M. Stevens was supported by Leverhulme Trust grant F/00577/1. Ishrat Jahan was supported by
EPSRC grant EP/E016839/1.

\providecommand{\noopsort}[1]{}\providecommand{\singleletter}[1]{#1}%

\end{document}